\documentstyle[aps,floats]{revtex}
\draft
\input epsf
\begin{document}
\hyphenation{arc-length fila-ment fila-ments geo-met-ry}
%
%
\twocolumn[\hsize\textwidth\columnwidth\hsize\csname
@twocolumnfalse\endcsname
\begin{flushright}
\baselineskip=14pt
{ FERMILAB--Pub--96/406-A}\\
{ {\it patt-sol/9611001\/} at {\it xyz.lanl.gov}}
\end{flushright}

\title{Vortex Dynamics in Dissipative Systems}
\author{Ola T\"ornkvist}
\address{NASA/Fermilab Astrophysics Center,
MS--209, P.O.\ Box 500, Batavia, IL~60510}
\author{Elsebeth Schr\"oder}
\address{Niels Bohr Institute,
Blegdamsvej 17, DK--2100 Copenhagen \O, Denmark}
\date{November 4, 1996}
\maketitle
\begin{abstract}
We derive the exact equation of motion for a vortex  in two-
and three-dimensional non-relativistic systems governed by
the Ginzburg-Landau equation with complex coefficients.
The velocity is given in terms of local gradients of
the magnitude and phase of the complex field and is exact
also for arbitrarily small inter-vortex distances.  The
results for vortices in
a superfluid or a superconductor are recovered.
\end{abstract}
%
]
%
\footnotetext{~\vspace{-.8cm}~}
\footnotetext{Electronic address: {\tt olat@fnal.gov}}
\footnotetext{Electronic address: {\tt schroder@nbi.dk}}
Vortices are found in a variety of physical systems. Accordingly,
the study of these intriguing collective excitations attracts
widespread attention among the physics community.
Examples of vortices often studied are hydrodynamic vortices,
vortices in superfluids, in superconductors and in nematic crystals,
and cosmic strings \cite{review,Batchelor}.
An important goal is to clarify the mechanisms by which
vortices are created and the details of their motion subject to local
interactions, such as crossing, merging and intercommutation,
as well as long-range forces.
These issues have recently been addressed in the context of relativistic
scalar field theories
\cite{BenYac1995,Ben-Yaacov}.

In this Letter we present an analytic derivation of the
exact equation of motion
for a vortex in a non-relativistic dissipative system.
The system we study is
one modelled by the extensively studied
 \cite{Hohenberg,Kuramoto,CGLref,Rica,Elphick}
Ginzburg--Landau equation with
complex coefficients (CGL)
\begin{equation}
{\textstyle\frac{d}{dt}}\,A=P(A,A^{\ast})A +b \nabla^2 A
\label{CGL}
\end{equation}
where $A=|A| \exp(iS)$ is a complex field, the function $P$ is
given by $P(A,A^{\ast})=\mu -a |A|^2$,
and $a,b,\mu\in {\bf C}$. By a suitable rescaling of time, space,
and $A$, the number of (real) adjustable parameters in the
coefficients of
eq.~(\ref{CGL}) may be brought down to two, as is often done.
However, we shall
keep eq.~(\ref{CGL}) unscaled for clarity. We study the equation
in two and three spatial dimensions.

The reason for selecting the CGL equation is two-fold.
Firstly, it is a relatively simple
partial differential equation; yet it exhibits the principal
features of more complicated oscillatory systems.  A prime example
of such systems are reaction-diffusion systems, such as the chemical
oscillatory Belousov-Zhabotinsky reaction\cite{BZ,Winfree}.

Secondly, the CGL equation contains a number of interesting
special cases. When $a$, $b$, and $\mu$ are purely
imaginary the CGL equation coincides
with the non-linear Schr\"{o}dinger equation. The latter equation
describes the quantum dynamics of superfluid $^4\!$He
and is known in that context as the Ginzburg-Pitaevskii-Gross equation
(GPG) \cite{GPGrefs}. Furthermore, by employing the Madelung
transformation \cite{Madelung} the non-linear
Schr\"{o}dinger equation also transforms into the hydrodynamic
equations for an inviscid fluid (the Euler equations).
In both cases $|A|^2$ corresponds to the (super)fluid mass density
and $\nabla S$ is proportional to the velocity of the (super)fluid.
We stress that the GPG case is special because it describes a
conservative system and the vortex motion can be derived from
a Lagrangian. The
CGL equation, on the other hand, describes a dissipative system and
one is compelled to pursue a direct derivation of the vortex
equation of motion, as we do here.

\begin{figure}
\hspace*{7mm}\epsfxsize=5cm \epsfbox[82 292 400 490]{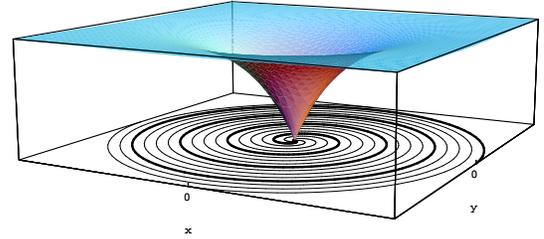}
\vspace*{.4cm}
\caption{One-armed spiral vortex of the CGL field $A=|A|e^{iS}$ in a
two-dimensional
system. The height of the surface depicts the magnitude $|A|$. The spiraling
curves are contour lines of the phase $S$ (isophase lines). The phase change
between two thin lines is $\pi/2$.}
\label{vortexfigure}
\end{figure}

Equation~(\ref{CGL}) permits solutions in which $A$
has phase singularities ({\em defects\/}).
In two space dimensions these are isolated points  around which the
phase $S$ changes
by multiples of $2\pi$. At the same points the magnitude
$|A|$ vanishes, so that the complex field $A$ remains single
valued, see figure~\ref{vortexfigure}.
In the vicinity of a defect
the phase is of the form
$S=\psi(r) +n(\varphi -\omega t)$ in polar coordinates ($r,\varphi$)
\cite{Hagan}.  For a constant phase $S$ this is the
equation  for
$|n|$-armed spirals
rotating at an angular frequency $\omega$.
In three dimensions the defects become
one-dimensional strings, or {\em filaments\/}, and the spirals
generalize
to scroll waves \cite{Winfree,Keener} which
look like sheets
wound around a filament. The filaments
may be closed or open (in which case
they end on the system boundary) and of arbitrary shape.
We shall call a solution with one defect or filament (in two or three
dimensions)
a spiral vortex, in analogy with the (non-spiral, $\psi'(r)=0$)
vortex solution of the GPG equation,
which describes the circulation of the superfluid around strings of
normal-phase fluid.
The integer $n$ is the winding number of the vortex
and is a topologically conserved quantity in two dimensions but not
in three. The {\em core\/} of the vortex is the
region where the magnitude $|A|$
deviates significantly from its asymptotic value,
see figure~\ref{vortexfigure}.

The evolution of a system with (spiral) vortices may
be described in terms of the motion of the defects, or filaments,
along with values
of the fields $|A|$ and $S$ at positions away from the
defects or filaments.
Such a separation into collective coordinates and field variables is
non-trivial, and the present work comprises the first exact treatment
of this kind for a dissipative system.
The motion of a vortex is affected by modifications in the field $A$
due to the presence of other vortices or system boundaries.
If the vortices are assumed to form a
dilute system, i.e.\  one where the
defects are well separated, the influence of variations in
the magnitude $|A|$ of the complex field may be neglected,
since $|A|$ will assume its asymptotic value
at distances much smaller than the inter-defect distance
\cite{incompress}.
Under this assumption, the interaction between
vortices can be described entirely by the
phase $S$. In this approximation Rica and Tirapegui \cite{Rica}
(and in a slightly different form also Ref.~\cite{Elphick})
have derived the equation of motion in two space dimensions
for the position of
the $k$th defect $\mbox{\boldmath $X$}_k(t)$ in terms of
the portion of the phase $S$ due to other defects,
$\theta^{(k)}(\mbox{\boldmath $x$})\equiv S - n_k \varphi_k$, where
$\tan\varphi_k=(y-Y_k)/(x-X_k)$. Their result  (for $|n_k|=1$ and
$b_{\scriptscriptstyle R}=1$, but here generalized to any value of
$n_k$ and $b_{\scriptscriptstyle R}$) is
\begin{equation} \label{ricaeq}
\dot{\mbox{\boldmath $X$}}_k\equiv
{{d\mbox{\boldmath $X$}_k}\over{dt}}=
2b_{\scriptscriptstyle I}\nabla\theta^{(k)} -
2b_{\scriptscriptstyle R}\frac{n_k}{|n_k|}
\hat{\mbox{\boldmath $z$}}
\times\nabla\theta^{(k)}
{\hspace{.15cm} {\rm ,}\hspace{-.15cm}}
\end{equation}
where $b_{\scriptscriptstyle R}\equiv {\rm Re}\,b$,
$b_{\scriptscriptstyle I}\equiv {\rm Im}\,b$, and
$\hat{\mbox{\boldmath $z$}} =\hat{\mbox{\boldmath $x$}} \times
\hat{\mbox{\boldmath $y$}} $ is normal to the plane.
The first term, proportional to the gradient, is that found by
Fetter \cite{Fetter} in the GPG limit corresponding to
$b_{\scriptscriptstyle R}=0$, $b_{\scriptscriptstyle I}=\hbar/(2m)$
and states that the vortex moves with the local superfluid velocity.
The second term is the perpendicular Peach-Koehler term \cite{Peach}
first found in this context by Kawasaki \cite{Kawasaki}.

When the system of spiral vortices cannot be
approximated by a dilute system
the expression (\ref{ricaeq}) for the defect velocity
is no longer valid but
will acquire additional terms.
We shall take a completely general approach in which
the amplitude $|A|$ is allowed to vary. This will
enable us to determine
the exact motion
of a defect also when another defect is located an
arbitrarily small distance away, i.e.\ even when the
vortex cores overlap.
It will also provide the exact motion
of a defect which is arbitrarily near a system boundary.
For filaments in a three-dimensional system
our treatment will furthermore correctly
incorporate interactions with other segments of
the same filament.

The corresponding problem for a relativistic scalar field theory was
solved by Ben-Ya'acov \cite{Ben-Yaacov}. His derivation
was based strictly on a covariant world-sheet formalism that cannot
be applied to a non-relativistic theory.
For the CGL equation one must therefore resort to other methods.

Let us consider the general motion of vortices in three space
dimensions.
The motion in a two-dimensional system can be found from the
three-dimensional problem as the special case of straight, aligned
vortices.

We may generalize eq.~(\ref{CGL}) by
admitting any continuous function $P(A,A^{\ast})$.
The details of $P$ do not enter the
derivation.
The field $A$ is
zero on a collection of one-dimensional strings  which
are the filaments.
Let the position of the filament $\Gamma$ of a vortex
be given at time $t$ by $\mbox{\boldmath $X$}(s,t)$, where
$s$ is the arclength coordinate along $\Gamma$.
We define a local coordinate system along the string as follows
\cite{diffgeo}.
At each point along the string
the unit tangent vector
$\mbox{\boldmath $T$}=\partial\mbox{\boldmath $X$}/\partial s$,
the unit normal vector $\mbox{\boldmath $N$}$, and the binormal
vector $\mbox{\boldmath $B$}=
\mbox{\boldmath $T$}\times\mbox{\boldmath $N$}$
form an orthonormal frame
so that any position
$\mbox{\boldmath $x$}$ in a neighborhood of the string
can be expressed as
$\mbox{\boldmath $x$}=\mbox{\boldmath $X$}(s,t)+
x\mbox{\boldmath $N$}(s,t)+y\mbox{\boldmath $B$}(s,t)$.
The coordinate representation $(s,x,y)$ is unique for
$x<1/\kappa$ but becomes singular when $x$ reaches or exceeds
the radius of curvature $1/\kappa$.

Along the string, the transport of the unit vectors is given by the
Frenet-Serret equations \cite{diffgeo}
\begin{equation} \label{frenet}
\frac{\partial\mbox{\boldmath $T$}}{\partial s} =
\kappa\mbox{\boldmath $N$}{ {\rm ,}\hspace{.20cm}}
\frac{\partial\mbox{\boldmath $N$}}{\partial s}=
-\kappa\mbox{\boldmath $T$} + \tau\mbox{\boldmath $B$}
{ {\rm ,}\hspace{.20cm}}
\frac{\partial\mbox{\boldmath $B$}}{\partial s}=
-\tau\mbox{\boldmath $N$}{ {\rm ,}}
\end{equation}
where $\kappa$ is the curvature and $\tau$ is the torsion of the
string.
Let us further introduce the
local polar coordinates $r$, $\varphi$ defined by $x=r \cos\varphi$,
$y=r \sin\varphi$. In terms of these coordinates, the gradient and
Laplacian take the forms
\begin{eqnarray}
\nabla &=& \mbox{\boldmath $T$} H +
\hat{\mbox{\boldmath $r$}}\frac{\partial\,}{\partial r} +
\hat{\mbox{\boldmath$\varphi$}}
\frac{1}{r}\frac{\partial\,}{\partial\varphi}\\
\nabla^2 &=& H^2 + \frac{\partial^2\,}{\partial r^2} +
\frac{1}{r}\frac{\partial\,}{\partial r}+
\frac{1}{r^2}\frac{\partial^2\,}{\partial\varphi^2} \nonumber \\
&&\mbox{} - \frac{\kappa}{1-\kappa r \cos\varphi}
\left(\cos\varphi \frac{\partial\,}{\partial r} -
\sin\varphi \frac{1}{r}\frac{\partial\,}{\partial\varphi}\right)\, ,
\label{laplace}
\end{eqnarray}
where
\begin{equation} \label{hdef}
H=\frac{1}{1-\kappa r\cos\varphi}\left(
\frac{\partial\,}{\partial s}-
\tau\frac{\partial\,}{\partial\varphi}
\right)\, .
\end{equation}

We now proceed to find the velocity
$\dot{\mbox{\boldmath $X$}}(s,t)$
of the filament $\Gamma$.  Because this
string of zeros of the function $A$ has no transverse extension
and is a feature of  a solution of
an underlying local field theory, its motion should be determined
from the behavior of the fields
$|A|$ and $S$ in an infinitesimal
neighborhood of the filament. It will be sufficient to
study the fields within a distance
$\varepsilon\ll \min(d,1/\kappa)$,
where $d$ is the shortest distance to another
string segment \cite{Comment}.
This condition ensures uniqueness of the coordinate
representation.

The phase field $S$ is multi-valued and satisfies
$S(s,r,\varphi+2\pi;t) - S(s,r,\varphi;t)=n\,2\pi$ for
$0<r<\varepsilon$.
Let us therefore split
$S=\chi +\theta$
in such a way that $\chi$
contains all multi-valued contributions
to the phase and depends on time only through the
position of the filament $\Gamma$.
For a straight (or two-dimensional) isolated
vortex one may choose $\chi=n\varphi$. A consistent description of
the multi-valued phase of an arbitrarily shaped
vortex filament requires, however, a global realization such as the
Biot-Savart integral,
\begin{equation}
\label{biot}
\nabla\chi = \frac{n}{2}\int_{\Gamma} d\mbox{\boldmath $X$}\times
\frac{\mbox{\boldmath $x$}-\mbox{\boldmath $X$}}
{|\mbox{\boldmath $x$}-\mbox{\boldmath $X$}|^3}
{\hspace{.15cm} {\rm .}\hspace{-.15cm}}
\end{equation}
This expression is known to contain
logarithmic divergencies as $r\to 0$,
as well as functions of the
azimuthal angle $\varphi$ that are multi-valued at $r=0$
\cite{Batchelor}.
We therefore absorb in $\chi$
any part of $S$ that is non-differentiable at
$r=0$.
Similarly, we may write $|A|=R w$,
where $\ln R$ depends on the filament position and
contains all contributions to $\ln |A|$ that are
non-differentiable at $r=0$.
For a straight isolated vortex one may choose
$R=r^{|n|}$.
Thus $\theta$ and $\ln w$ are differentiable and it follows that the
time derivatives
$\dot{\theta}$ and $\dot{w}$ are finite for
$r<\varepsilon$.
We remark that the choice of $\chi$ and $R$ is not unique, since
$S$ and $|A|$ are
invariant under two independent local symmetries
\begin{equation}
\label{symmetry}
\chi \to \chi+\delta,~
\theta\to\theta-\delta\quad;\quad
R\to R f, ~w\to wf^{-1}{\hspace{.15cm} {\rm ,}\hspace{-.15cm}}
\end{equation}
where $\delta$ and $\ln f$ are differentiable.

With these definitions
the real and imaginary parts of equation
(\ref{CGL}) lead to the two equations
\begin{eqnarray} \label{Reqans}
{\textstyle\frac{d}{dt}}\ (\ln R + \ln w) &= &
{\rm Re}(P)+ b_{\scriptscriptstyle R} Q_1
-b_{\scriptscriptstyle I} Q_2
{\hspace{.15cm} {\rm ,}\hspace{-.15cm}}\\*
{\textstyle\frac{d}{dt}}\ (\theta+  \chi) &=& {\rm Im}(P) +
b_{\scriptscriptstyle I} Q_1 + b_{\scriptscriptstyle R} Q_2
{\hspace{.15cm} {\rm ,}\hspace{-.15cm}} \label{Seqans}
\end{eqnarray}
where
\begin{eqnarray}
Q_1 &=&  \nabla^2 \ln R + \nabla^2 \ln w \nonumber\\*
&&\mbox{}+(\nabla\ln R + \nabla\ln w)^2
 - (\nabla\chi+\nabla\theta)^2{\hspace{.15cm}
{\rm ,}\hspace{-.15cm}}
\nonumber \\*
Q_2 &=&  \nabla^2\chi +\nabla^2\theta +
2 (\nabla \ln R+\nabla \ln w)\cdot (\nabla\chi+\nabla\theta).\nonumber
\end{eqnarray}
The time derivative $d/dt$ in eqs.~(\ref{Reqans}) and (\ref{Seqans}),
which is to
be evaluated in the lab frame, is related to the time derivative
$\partial/\partial t$
in the moving reference frame of the local segment of the
filament by
$d/dt = - \dot{\mbox{\boldmath $X$}}
\cdot\nabla + \partial/\partial t$.

In order to include logarithmic divergencies as well as
multi-valuedness as $r\to 0$,
we write $\nabla\chi= f_1 \hat{\mbox{\boldmath $r$}} +
(n/r + f_2)  \hat{\mbox{\boldmath $\varphi$}} + \lambda_1
\mbox{\boldmath $T$}$
 and
$\nabla\ln R=(|n|/r + f_3) \hat{\mbox{\boldmath $r$}} +
f_4 \hat{\mbox{\boldmath $\varphi$}}+ \lambda_2
\mbox{\boldmath $T$}$,
where
\begin{equation}
\label{fexp}
f_i(r,\varphi,s,t) = g_i(\varphi,s,t) + h_i(\varphi,s,t)
\ln{\kappa r}
+ {\cal O}(r)
\end{equation}
and ${\cal O}(r)$ denotes any terms that vanish as $r\to 0$.
It can be easily confirmed from this equation (as well as argued on
general grounds) that $\partial\chi/\partial t$,
$\partial\chi/\partial s$, $\partial(\ln R)/\partial t$,
$\partial(\ln R)/\partial s$, $\lambda_1$ and $\lambda_2$
 have well-defined finite limits as
$r\to 0$. We require that $\nabla\chi$ and $\nabla\ln R$ be integrable,
and that they satisfy the following condition near the filament:
\begin{equation}
\label{ortho}
\nabla\chi - \frac{n}{|n|}\mbox{\boldmath $T$}\times\nabla\ln R=
\mbox{\boldmath $C$}(s,t)+{\cal O}(r)
{\hspace{.15cm} {\rm .}\hspace{-.15cm}}
\end{equation}
The arbitrary vector $\mbox{\boldmath $C$}$
corresponds to a choice of gauge in eq.~(\ref{symmetry}).
In the {\em symmetric gauge\/}
 $R=r^{|n|}$, $\chi=n\varphi$  for a straight (or two-dimensional)
isolated vortex
we have $\mbox{\boldmath $C$}={\bf 0}$.

Since $\theta$ and $\ln w$ are differentiable,
the singularities of  $\nabla\chi$ and $\nabla\ln R$ at $r=0$ must
satisfy eqs.~(\ref{Reqans}) and
(\ref{Seqans}) order by order.
This last condition together with eq.~(\ref{ortho}),
leads to the coupled non-linear system
\begin{eqnarray}
\label{singeqs}
\nabla\ln R\cdot\mbox{\boldmath $u$} +
b_{\scriptscriptstyle R} q_1 - b_{\scriptscriptstyle I} q_2 &=&
{\rm regular} \nonumber\\*
\nabla\chi\cdot\mbox{\boldmath $u$} +
b_{\scriptscriptstyle I} q_1 +b_{\scriptscriptstyle R} q_2 &=&
{\rm regular}
{\hspace{.15cm} {\rm ,}\hspace{-.15cm}}
 \end{eqnarray}
where
$q_1=\nabla^2\ln R + (\nabla\ln R)^2 - (\nabla\chi)^2$,
$q_2=\nabla^2\chi + 2\nabla\ln R\cdot\nabla\chi$ and
$\mbox{\boldmath $u$}=\dot{\mbox{\boldmath $X$}} +
2b_{\scriptscriptstyle R} (\nabla\ln w + \frac{n}{|n|}
\mbox{\boldmath $T$}\times\nabla\theta) -
2b_{\scriptscriptstyle I} (\nabla\theta - \frac{n}{|n|}
\mbox{\boldmath $T$}\times\nabla\ln w)$.

Cancellation of  terms of order $r^{-1}$ in eq.~(\ref{singeqs})
leads to two equations for the perpendicular
components of $\mbox{\boldmath $u$}$. The integrability
condition provides four first-order differential equations
relating the functions $g_i$ and $h_i$,
and together with four algebraic relations resulting from
eq.~(\ref{ortho}) the system
can be solved in terms of four constants of integration.
The perpendicular components of
$\mbox{\boldmath $u$}$ are then uniquely determined in terms of
\mbox{\boldmath $C$}.
Furthermore, the singular terms of order $r^{-1}\ln\kappa r $ in
eq.~(\ref{singeqs})
cancel.
It is always possible to set
the tangential velocity, which is void of physical meaning, to
zero by a time-dependent
reparametrization $s \to s(t)$. In the language of relativistic
string theory, this is referred to as world-sheet reparametrization
invariance. The exact result for the velocity of the vortex filament
is
\begin{eqnarray} \label{veq}
{\dot{\mbox{\boldmath $X$}}}&=&
{b_{\scriptscriptstyle I} \left(\kappa {{n}\over{|n|}}
\mbox{\boldmath $B$} +
2(\nabla_\perp{\theta}+\mbox{\boldmath $C_\perp$})-
2{{n}\over{|n|}}\mbox{\boldmath$T$}\times
\nabla\ln {w} \right)}\nonumber\\*
&+&{b_{\scriptscriptstyle R}\left(\kappa \mbox{\boldmath $N$} -
2\nabla_\perp\ln w -
2\frac{n}{|n|}\mbox{\boldmath $T$}\times (\nabla\theta+
\mbox{\boldmath $C$})
\right)}{\hspace{.15cm} {\rm ,}\hspace{-.15cm}}
\end{eqnarray}
where $(~)_\perp = - \mbox{\boldmath $T$}\times
[\mbox{\boldmath $T$} \times (~)]$
and the fields on the right-hand side are to be evaluated at
the filament position $\mbox{\boldmath $X$}(s,t)$.
The exact two-dimensional result is obtained as $\kappa\to 0$.

The value of  $\dot{\mbox{\boldmath $X$}}$ is
independent of the choice of gauge
for $R$ and $\chi$. Indeed, substituting  $\mbox{\boldmath $C$}$ from
eq.~(\ref{ortho}) into eq.~(\ref{veq}) we obtain
the manifestly invariant expression
\begin{eqnarray} \label{vA3d}
{\dot{\mbox{\boldmath $X$}}}&=&
\lim_{r\to 0}
\left[
{b_{\scriptscriptstyle I}\!\!\left(\kappa
{{n}\over{|n|}}\mbox{\boldmath $B$} +
2\nabla_\perp S - 2{{n}\over{|n|}} \mbox{\boldmath $T$}\times
\nabla\ln |A| \right)}\right.\nonumber\\*
&&~~~~~~+\left.{b_{\scriptscriptstyle R}\!\!\left(\kappa
\mbox{\boldmath $N$} -
2 \nabla_\perp\ln |A| - 2{{n}\over{|n|}} \mbox{\boldmath $T$}\times
\nabla S \right)}\right].{\hspace{-0.1cm}}
\end{eqnarray}
in which the filament velocity is written in terms of gradients of the
magnitude and
phase of the original complex field $A$. Let us define the complex
velocity
$\dot{Z}\equiv
(\mbox{\boldmath $N$}+i \mbox{\boldmath $B$})\cdot
\dot{\mbox{\boldmath $X$}}$
and express the derivatives in eq.~(\ref{vA3d}) in terms of
$z\equiv x+i y$ and its conjugate $z^{\ast}$. Then a quite
beautiful result emerges:
\begin{equation} \label{vanal3d}
\left\{ \begin{array}{ll}
\dot{Z} = b\left(
- 4 \frac{\partial~}{\partial z^{\ast}} \ln A(z,z^{\ast}) +
 \kappa\right) &,\ n\geq 1\\*[2ex]
\dot{Z}{}^{\ast} =
b \left(-4 \frac{\partial~}{\partial z}\ln A(z,z^{\ast})+\kappa\right)
&,\ n\leq-1
{\hspace{.15cm} {\rm ,}\hspace{-.15cm}}\end{array}\right.
\end{equation}
where the right-hand side is to be evaluated at
$z\!=\!z^{\ast}\!=\!0$.

The results are to be interpreted as follows: The velocity of the
central filament of a vortex gets contributions from the curvature $\kappa$
of the filament and from local gradients of the magnitude $|A|$ and
phase $S$ of the complex field.  A cylindrically symmetric solution
$A=\rho(r)  \exp[i (\psi(r) + n(\varphi - \omega t))]$,
for which $\rho=|A|\sim r^{|n|}$ and $\psi'(0)=0$,
contributes nothing to the velocity and corresponds to a straight
(or two-dimensional) isolated vortex at rest with respect to the lab
frame. Non-zero gradient contributions appear as a result of deviations
from cylindrical symmetry in $|A|$ and $S$. In a symmetric gauge with
$\mbox{\boldmath $C$}={\bf 0}$, these deviations are represented by
$w$ and $\theta$. The asymmetries arise from
the presence of other vortices, system boundaries, or  (in three
dimensions) other segments of the same filament, causing the vortex to move.

In the $\mbox{\boldmath $C$}={\bf 0}$ gauge the expression
(\ref{veq}) reproduces a variety of results obtained previously for
special cases.  For $\kappa=0$ and $\nabla\ln w\approx 0$ it
reduces to
eq.~(\ref{ricaeq}) corresponding to a two-dimensional dilute system
\cite{Rica}.
In the GPG limit $b_{\scriptscriptstyle R}=0$
the expression (\ref{veq}) coincides with that derived by  Lee \cite{Lee},
who used a different method to find the velocity.
For $b_{\scriptscriptstyle I}=0$, eq.~(\ref{CGL})
describes the non-linear diffusion of two fluid components with
identical diffusion constants. In this limit the contribution to
${\dot{\mbox{\boldmath $X$}}}$ from curvature,
$b_{\scriptscriptstyle R}\kappa\mbox{\boldmath $N$}$, agrees
with the result of Ref.~\cite{Keener}.

The expressions (\ref{veq})--(\ref{vanal3d}) for the
velocity are exact also for an arbitrarily small distance between
filaments. This makes the formulation well suited for
theoretical or numerical investigations of local vortex
interactions, such as crossing, merging and intercommutation, in which
the vortex cores overlap \cite{BenYac1995,Dziarmaga}.
We caution that the GPG equation
does not provide a realistic model
for the core of a superfluid vortex, since there the
core width is comparable
to interatomic distances. For magnetic flux vortices
in a superconductor, however,
the core width is much larger and a classical description is justified.
Such vortices are solutions of eq.~(\ref{CGL})
with the substitution $\nabla \to \nabla +
2ie\mbox{\boldmath $A$}/(\hbar c)$,
where $\mbox{\boldmath $A$}$ is the vector potential
and $2e$ is the charge of a Cooper pair.
The corresponding filament velocity is easily obtained by adding
$2e\mbox{\boldmath $A$}/(\hbar c)$ to $\nabla\theta$ in
eq.~(\ref{veq}) or to $\nabla S$ in eq.~(\ref{vA3d}) \cite{Lee}.

In summary, we have derived the exact equation
of motion for a vortex in a large class of models
of a non-relativistic complex field
described by the complex Ginzburg--Landau equation (\ref{CGL}) with
an arbitrary, continuous function $P(A,A^{\ast})$. The velocity is
expressed in terms of local gradients of  the magnitude and phase of the
complex field $A$. The result agrees
with that of Ref.~\cite{Rica} (our eq.~\ref{ricaeq})
in the case of a dilute two-dimensional
system of vortices, but for the general non-dilute
case in two and three dimensions we find additional contributions to the
velocity corresponding to the asymmetry of the magnitude $|A|$ around the
vortex.

We are grateful to E. Copeland for crucial remarks relating to the
three-dimensional case and to T. Bohr for useful discussions and for
pointing us in the direction of Ref.~\cite{Rica}.
Support for O.T. was provided by DOE and NASA under Grant NAG5-2788
and by the Swedish Natural Science Research Council (NFR), for E.S.
by the Danish Natural Science Research Council.


\end{document}